# A FIBRE SMART DISPLACEMENT BASED (FSDB) BEAM ELEMENT FOR THE NONLINEAR ANALYSIS OF R/C MEMBERS

**Bartolomeo Pantò[1], Davide Rapicavoli[2], Salvatore Caddemi[3], Ivo Caliò[3]**

[1]Department of Civil Engineering and Architecture, University of Catania, Catania, Italy

Beam finite elements for non linear plastic analysis of beam-like structures are formulated according to Displacement Based (DB) or Force Based (FB) approaches. DB formulations rely on modelling the displacement field by means of displacement shape functions. Despite the greater simplicity of DB over FB approaches, the latter provide more accurate responses requiring a coarser mesh. To fill the existent gap between the two approaches the improvement of the DB formulation without the introduction of mesh refinement is needed. To this aim the authors recently provided a contribution to the improvement of the DB approach by proposing new enriched adaptive displacement shape functions leading to the Smart Displacement Based (SDB) beam element.

In this paper the SDB element is extended to include the axial force-bending moment interaction, crucial for the analysis of r/c cross sections. The proposed extension requires the formulation of discontinuous axial displacement shape functions dependent on the diffusion of plastic deformations. The stiffness matrix of the extended smart element is provided explicitly and shown to be dependent on the displacement shape functions updating. The axial force-bending moment interaction is approached by means of a fibre discretisation of the r/c cross section. The extended element, addressed as Fibre Smart Displacement Based (FSDB) beam element, is shown to be accurate and furthermore accompanied by the proposal of an optional procedure to verify strong equilibrium of the axial force along the beam element, which is usually not accomplished by DB beam elements. Given a fixed mesh discretisation the performance of the FSDB beam element is compared with the DB approach to show the better accuracy of the proposed element.

**INTRODUCTION**

The non linear behaviour of reinforced concrete (r/c) frame structures suffering pre- and post-peak damage, up to collapse, is reasonably described numerically by means of the adoption mono-dimensional beam finite elements undergoing plastic deformations. Non linear plastic beam elements cannot be considered capable of a detailed description of the phenomenological behaviour of a r/c element if compared to a two or three dimensional finite element, however, they represent a good compromise between computational efficiency and accuracy for the numerical analysis of rather complex frame structures. The development of plastic deformations along beam elements is studied in the current practise by means of two different strategies proposed in the literature: a concentrated/lumpedplasticity approach and a distributed plasticity model. A comprehensive analysis and critical discussion of the two approaches is reported in [Almeida et a. 2016] together with an extensive literature therein contained. According to the first approach the plastic behaviour is considered lumped at pre-established beam cross sections (plastic hinges); however, the concept of equivalent plastic hinge length is introduced as a fictitious semi-empirical measure to account for various basic hypotheses and also match experimental results. The greatest advantage of the concentrated plasticity approach consists in its simplicity and reduced computational cost. The diffusion of plasticity along a region of the beam member, that led to the alternative definition of the plastic hinge length in the concentrated plasticity approach, is, on the contrary, a phenomenon partially addressed by the distributed plasticity models. In the latter models the spread of plasticity is allowed throughout chosen control cross sections along the beam, representative of a beam portion, where the non linear plastic constitutive equations are step-by-step integrated and for the latter reason called integration points. The analysis of frame structures by means of distributed plasticity models, despite the introduction of a sort of algorithmic complexity, is widely used particularly in view of the need of accurate results and a better description of real phenomena.

Distributed plasticity models can be distinguished into Displacement Based (DB) or Force Based (FB) approaches according to the finite element formulation adopted. DB approaches are based on assumptions related to the displacement fields along the element. Usually cubic polynomials for the transversal displacement field and linear functions for the axial displacement are assumed corresponding to a linear curvature and constant axial strain along the element [Crisfield 1986, Crisfield 1991, Crisfield 1997, Zienkiewicz and Taylor 1987, Zienkiewicz and Taylor 2000, Bathe 1996]. On the contrary, FB approaches start from the description of the internal force field, obtained by the exact integration of the governing differential equations of the element, irrespective of the occurrence of plastic deformation [Ciampi and Carlesimo 1986, Carol and Murcia 1989]. Despite the greater simplicity and wider code diffusion of DB over FB approaches, the former are based on displacement shape functions, which are exact only for the linear elastic behaviour, while the second propose force shape functions are not affected by the occurrence of non-linearities. For the latter reason FB approaches, somehow affected by difficulties related to the availability of constitutive laws in terms of flexibility or else by the triggering of additional internal iterations, have been widely studied and improved in the recent literature [Taucer et al. 1991, Spacone et al. 1992, Spacone et al. 1996, Neuenhofer and Filippou 1997, Coleman and Spacone 2001, Scott and Fenves 2006, Adessi and Ciampi 2007, Scott and Hamutçuoğlu 2008, Almeida et al. 2012, Saritas and Soydas 2012, Adessi et al. 2015], by affirming a supremacy in terms of accuracy of the results.

On the other hand, since FB beam finite elements do not rely on displacement shape functions, they are not, for their nature, dedicated to the reconstruction of the displacement field during the analysis which requires a double integration procedure for the above purpose [Neuenhofer and Filippou 1998].

On the other hand, numerous studies, aiming at achieving more accurate results with the use of DB elements by introducing a dramatic mesh refinement and huge computational costs, can be found in the literature [Calabrese et al. 2010]. Improvements in the mesh refinementbased DB procedures have been achieved in [Izzudin and Elnashai 1993a, Izzudin and Elnashai 1993b, Izzudin et al. 1994, Karayannis et al. 1994, Izzuddin et al. 2002] where each structural member (upgraded with quartic shape functions) is checked upon occurrences of plastic deformations and, if the case, subjected to restricted automatic subdivision into elasto-plastic elements with cubic shape functions.

The latter adaptive mesh refinement practise, providing better accuracy and efficiency of the DB approach, has been implemented in a nonlinear analysis program ADAPTIC [Izzudin 1989] which can be efficaciously adopted to perform comparisons the with FB approach both in terms of accuracy and computational demand.

A study aimed at evaluating the performance of DB vs FB approaches accounting for both mesh discretisation and number of integration points is reported in detail in [Almeida et al. 2016] with regard to cases of hardening and softening behaviour.

Within the context of distributed plasticity along the beam element length, in conjunction with the development of the FB approach, a refined discretisation of the cross sections, placed at the integration points, into small cells or strips which follow uni-axial non linear constitutive laws has been introduced [Spacone et al. 1996a, Spacone et al. 1996b, Ceresa et al. 2009, Li et al. 2018]. By doing so each cell is representative of a material fibre and detailed diffusion of plasticity along the cross section can be described, by considering an appropriate number an postions of cross control sections, and any calibration of the bi-axial constitutive law is avoided. The cross section discretisation is usually referred to as fibre model and it is conveniently adopted for the analysis of composite materials such as r/c cross sections [Kaba and Mahin 1984, Zeris and Mahin 1988].

Given the above scenario more work should be devoted towards better performances of the DB beam elements. In fact, the classical polynomials adopted to formulate the displacement shape functions do not lend themselves to properly describe the displacement field during the inelastic analysis. Hence formulations of new conceived displacement shape functions, able to follow the development of plastic deformations with neither degree of freedom nor shape function order increments, is auspicable.

To the aim of improving the performance of the DB approach avoiding any cumbersome mesh refinement, the authors recently formulated new enriched adaptive displacement shape functions able to update in accordance to the diffusion of the plastic deformations during the analysis. With respect to the classical Hermite cubic polynomials, the proposed shape functions contain additional terms (formulated by means of generalised functions) that update during the inelastic analysis being dependent on stiffness decay of the cross sections according to a reference stepped beam model as diffusion of plastic deformations occurs. The resulting element, formulated in a pure flexural state and according to a sectional approach for homogeneous sections, was called Smart Displacement Based (SDB) beam element [Pantò et al. 2017]. Given the efficacy of the smart displacement shape functions introduced in [Pantò et al. 2017], within the same framework, however in the context of concentrated plasticity, shape functions endowed with additional terms, able to capture the occurrence of plastic hinges during the analysis at any beam cross section, have been adopted in [Hajidehi et al. 2018, Spada et al. 2018] to improve the DB formulation of concentrate plasticity inelastic beams.

In this paper the SDB beam element is improved and some of the original limitations removed. Precisely, a formulation to include the axial force-bending moment interaction, particularly crucial for the analysis of r/c cross sections is presented. The proposed formulation requires the introduction of additional degrees of freedom to describe the axial behaviour and the definition of discontinuous axial displacement shape functions based on a stepped distribution of the axial stiffness degradation dependent on the diffusion of plastic deformations. The stiffness matrix of the extended smart element is provided explicitly and shown to be dependent on the displacement shape functions updating. To avoid a bi-axial integration of the constitutive equations over the cross section the axial force-bending moment interaction is approached by means of a fibre discretisation suitable for r/c cross sections. The extended element is named in this work Fibre Smart Displacement Based (FSDB) beam element. The intention of the authors has been guided by the purpose of delivering a new DB beam element that does not require sub-discretisation of each member in a r/c frame structure. The proposed FSDB beam element is furthermore accompanied by the proposal of a procedure to verify strong equilibrium of the axial force along the beam element which is usually not accomplished by DB beam elements. Precisely, the variation of axial load along the integration points is iteratively corrected by the imposition of explicitly formulated fictitious axial strain distributions until it vanishes. The resulting element is axially equilibrated providing further improvement of the already performing smart displacement shape functions.

Given a fixed mesh discretisation the performance of the FSDB beam element is compared with the DB approach to show the better accuracy of the proposed element.

## THE SMART DISPLACEMENT SHAPE FUNCTIONS (SDSFS) FOR TRANSVERSAL AND AXIAL DISPLACEMENTS

Occurrence of plastic deformations along portions of a beam element is a phenomenon responsible of the stiffness decay of the beam cross sections due to exceedance of the stress level over the elastic range. A close description of the current beam stiffness in the analytical model would lead to governing differential equations with variable coefficients. Even then, the correct nonlinear spatial evolution of the stiffness during the time step plastic analysis cannot be followed and is rather conducted by means of approximated models. The starting point of the latter models is represented by the beam models with along axis variable stiffness. Since the stiffness variability is not a priori known, nonlinear models to follow the diffusion of plasticity usually rely on the hypothesis that in the plastic range beams behave according to a stepped variation of the plastic stiffness. Even though attempts to describe continuous variations of the beam plastic stiffness along the axis can be made, the stepped beam model remains quite appealing in view of its consistency with the Gauss integration scheme usually adopted to calculate the beam tangent stiffness matrix. In fact, the well known Gauss weights, coupled to the integrand function evaluated at the so-called Gauss points, represent effectively the portion of the beam where the plastic stiffness is assumed uniformly distributed.

The above discussion convinced, in the past, some of these authors to devote attention to the study of beams with stepped variations of the bending stiffness [Biondi and Caddemi 2005, Biondi and Caddemi 2007] and also to formulate linear two node finite elements with different types of singularities also by adopting the classical Timoshenko theory to account for the shear deformations [Caddemi et al. 2013a, Caddemi et al. 2013b]. The governing equations of the flexural behaviour of stepped beams has been successfully applied for the plastic analysis of beam elements [Pantò et al. 2017], however, in the latter work plastic deformations in axial direction have not been accounted for and the effect of the related axial stiffness decay has not been taken into account. Due to the above limitation the theory therein presented is not applicable to frame structures. For the latter reason, in this section the formulation of the stepped beam model both in the axial and transversal direction is presented and the relevant closed form solution, put in a form suitable to be exploited for a more complete inelastic beam element for diffused plasticity, is derived.

An $x, z$ plane model for axial $E(x)A(x)$ and flexural $E(x)I(x)$ stiffnesses of beam cross sections with stepped variations along the beam axis is depicted in Figure 1, where abscissa $x$ spans from $0$ to the length $L$ of the beam. $E(x), A(x), I(x)$ represent the Young modulus, the area and the moment of inertia of the cross section at abscissa $x$, respectively. The beam model under consideration is hence characterised by $n$ segments with abrupt stiffness changes and can be formulated by making use of the well know Heaviside (unit step) generalised function $U(x-x_i)$, as follows:

$$E(x)A(x) = E_o A_o \left[ 1 - \sum_{i=1}^{n} (\beta_{x,i} - \beta_{x,i-1}) U(x-x_i) \right]$$

$$E(x)I(x) = E_o I_o \left[ 1 - \sum_{i=1}^{n} (\beta_{z,i} - \beta_{z,i-1}) U(x-x_i) \right] \tag{1}$$

where $x_i$ indicates the abscissa along the beam axis where the $i$-th cross section change occurs, and $\beta_{x,i} - \beta_{x,i-1}$ and $\beta_{z,i} - \beta_{z,i-1}$ denote the relevant axial and bending stiffness abrupt variations with respect to the reference values $E_o A_o$ and $E_o I_o$, respectively, where $E_o, A_o, I_o$ represent the reference values of the Young modulus, the area and the moment of inertia of the cross section, respectively. The stepped beam model depicted in Figure 1 and described analytically by the model introduced in Equation (1) implies that the beam is composed of $n$ segments with axial stiffness $E_i A_i$, $i=1,\ldots,n$, assuming $\beta_{x,i} = \dfrac{E_o A_o - E_i A_i}{E_o A_o}$, and flexural stiffness $E_i I_i$, $i=1,\ldots,n$, where $\beta_{z,i} = \dfrac{E_o I_o - E_i I_i}{E_o I_o}$.

Making use of the model in Equation (1) into the governing equations of the variable cross section Euler-Bernoulli beam subjected to a static axial $p_x(x)$ and transversal load $p_z(x)$ distribution leads to the following generalized differential equations:

$$E_o A_o \left\{ \left[ 1 - \sum_{i=1}^{n} (\beta_{x,i} - \beta_{x,i-1}) U(x-x_i) \right] u_x'(x) \right\}' = -p_x(x)$$

$$E_o I_o \left\{ \left[ 1 - \sum_{i=1}^{n} (\beta_{z,i} - \beta_{z,i-1}) U(x-x_i) \right] u_z''(x) \right\}'' = p_z(x) \tag{2}$$

where the apex indicates the differentiation with respect to $x$ and $u_x(x), u_z(x)$ are the axial and the transversal deflection functions.

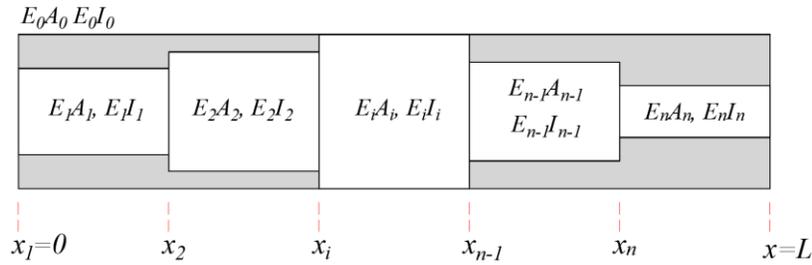

**Figure 1 Equivalent stepped axial-flexural beam.**

Integration of Equation (2) according to the generalised function integration rules [33-39] leads to:

$$u_x'(x) = -\dfrac{1}{E_o A_o \left[ 1 - \sum_{i=1}^{n} (\beta_{x,i} - \beta_{x,i-1}) U(x-x_i) \right]} \left[ p_x^{[1]}(x) + d_1 \right]$$

$$u_z''(x) = \dfrac{1}{E_o I_o \left[ 1 - \sum_{j=1}^{n} (\beta_{z,i} - \beta_{z,i-1}) U(x-x_i) \right]} \left[ p_z^{[2]}(x) + b_1 x + b_2 \right] \tag{3}$$

Where $d_1, b_1, b_2$ are integration constants and $p_x^{[k]}(x), p_z^{[k]}(x)$ indicate the $k$-th primitive functions of the relevant external load distributions $p_x(x), p_z(x)$, respectively.

In view of the properties of the Heaviside generalised function [Bremermann and Durand 1961, Colombeau 1984, Guelfand and Chilov 1972, Hoskins 1979, Kanwal 1983, Lighthill 1958, Zemanian 1965], Equation (3) can also be written as follows:

$$u_x^I(x) = -\left[1 + \sum_{i=1}^{n} \beta_{x,i}^* U(x-x_i)\right]\left[\frac{p_x^{[1]}(x)}{E_o A_o} + \frac{d_1}{E_o A_o}\right]$$

$$u_z^{II}(\xi) = \left[1 + \sum_{i=1}^{n} \beta_{z,i}^* U(x-x_i)\right]\left[\frac{p_z^{[2]}(x)}{E_o I_o} + \frac{b_1}{E_o I_o}x + \frac{b_2}{E_o I_o}\right]$$

(4)

Where, in order to obtain a more compact notation, the following new parameters $\beta_{x,i}^*, \beta_{z,i}^*$ have been defined:

$$\beta_{x,i}^* = \frac{\beta_{x,i}}{1-\beta_{x,i}} - \frac{\beta_{x,i-1}}{1-\beta_{x,i-1}} \quad , \quad \beta_{z,i}^* = \frac{\beta_{z,i}}{1-\beta_{z,i}} - \frac{\beta_{z,i-1}}{1-\beta_{z,i-1}}$$

(5)

Integration of Equation (4), in view of the integration rules of the distributions and after simple algebra, leads to the following explicit expressions for the axial displacement and the transversal deflection functions $u_x(x), u_z(x)$ respectively:

$$u_x(x) = a_1 + a_2 g_2(x) + g_3(x)$$

$$u_z(x) = c_1 + c_2 x + c_3 f_3(x) + c_4 f_4(x) + f_5(x)$$

(6)

where the functions $g_2(x), g_3(x), f_3(x), f_4(x), f_5(x)$ are defined as follows:

$$g_2(x) = -x - \sum_{i=1}^{n} \beta_{x,i}^*(x-x_i)U(x-x_i)$$

$$g_3(x) = -\frac{p_x^{[2]}(x)}{E_o A_o} - \sum_{i=1}^{n} \frac{\beta_{x,i}^*}{E_o A_o}\left[p_x^{[2]}(x) - p_x^{[2]}(x_i)\right]U(x-x_i)$$

$$f_3(x) = \left[x^2 + \sum_{j=1}^{n} \beta_{z,i}^*(x-x_i)^2 U(x-x_i)\right]$$

(7)

$$f_4(x) = x^3 + \sum_{j=1}^{n} \beta_{z,i}^*(x^3 - 3x_i^2 x + 2x_i^3)U(x-x_i)$$

$$f_5(x) = \frac{p_z^{[4]}(x)}{E_o I_o} + \sum_{i=1}^{n} \frac{\beta_{z,i}^*}{E_o I_o}\left[p_z^{[4]}(x) - p_z^{[4]}(x_i)\right]U(x-x_i) - \sum_{i=1}^{n} \beta_{z,i}^* p_z^{[3]}(x_i)(x-x_i)U(x-x_i)$$

In Eqs (7) some integration constants have been re-defined as $a_2 = d_1/(E_o A_o)$ $c_3 = b_2/(2E_o I_o), c_4 = b_1/(6E_o I_o)$, and the additional constants $a_1, c_1, c_2$ have been introduced.

Closed form expression of the normalised rotation $\varphi_y(x)$, curvature $\chi_y(x)$, axial force $N(x)$, bending moment $M_y(x)$ and shear force $T_z(x)$ functions are straightforwardly related to the expressions in Equation (6) by exploiting the standard expressions of the Euler-Bernoulli beam model, and in view of the adopted model in Equation (1), as follows:

$$\varphi_y(x) = -u_z^I(x) \quad , \quad \chi_y(x) = -u_z^{II}(x)$$

$$N(x) = E_o A_o \left[1 - \sum_{i=1}^{n}(\beta_{x,i} - \beta_{x,i-1})U(x-x_i)\right]u_x^I(x)$$

$$M_y(x) = E_o I_o \left[1 - \sum_{i=1}^{n}(\beta_{z,i} - \beta_{x,i-1})U(x-x_i)\right]\chi_y(x)$$

$$T_z(x) = -E_o I_o \left[1 - \sum_{i=1}^{n}(\beta_{z,i} - \beta_{x,i-1})U(x-x_i)\right]u_z^{III}(x)$$

(8)

In view of the definitions reported in Equation (7), it has to be remarked that the axial displacement $u_x(x)$ and the transversal deflection $u_z(x)$ functions, provided by Equation (6), are continuous functions, despite the presence of the Heaviside generalised function, together with the rotation $\varphi_y(x)$. On the contrary, discontinuities are recovered at cross sections $x_i$ in the first derivative of the axial displacement $u'_x(x)$ and in the curvature function $\chi_y(x)$ due to the axial and flexural stiffness abrupt changes of the stepped beam under study. Nevertheless, application of the relations in Equation (8) leads to continuous axial force $N(x)$, bending moment $M_y(x)$ and shear force $T_z(x)$ functions.

Equation (6), where the integration constants $a_1, a_2, c_1, c_2, c_3, c_4$ are to be determined by imposing the relevant boundary conditions, represents the explicit solution of the multi-stepped Euler-Bernoulli beam subjected to any external transversal load. It is worth noting that continuous, discontinuous and singular (concentrated load) distribution laws can be accommodated in Equation (6) by considering appropriate expressions for the terms related to the external load.

The stepped beam model presented in this section allows the formulation of a specific beam finite element. When variations of axial and flexural stiffness along the beam axis represent the stiffness decay of the beam cross sections, due to exceedance of the stress level over the elastic range, they may change during the evolutionary analysis giving rise to the adaptive (i.e. 'smart') element described in the next section.

## THE SMART DISPLACEMENT BASED (SDB) BEAM ELEMENT

The closed form solutions in terms of axial and transversal displacement of a stepped beam presented in the previous section can be conveniently adopted to formulate an extension of the Smart Displacement Based (SDB) beam element as formulated in [Pantò et al. 2017] for the analysis of inelastic beams under flexural behaviour only.

In this section a DB beam finite element able to account for the flexural and axial behaviour is proposed. The beam element is "smart" in the sense that the shape functions change accordingly to the inelastic response of the beam. Namely, the displacement shape functions here formulated, for both the axial and transversal displacements, contain additional terms, with respect to the usual linear and cubic polynomials, that do not remain constant but are subjected to update as axial and flexural stiffness decay occur due to the onset of plastic deformations.

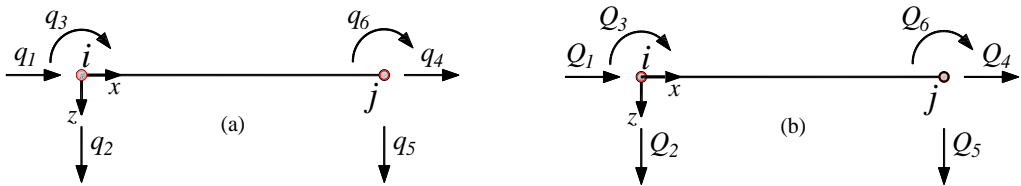

**Figure 2**   Nodal degrees of freedom (a) and dual forces (b) of the element.

The new beam element, connecting joints $i$ and $j$, is defined in the $x,z$ plane as shown in Figure 2 where the nodal displacements, given by the axial and transversal displacements and rotations at nodes $i$ and $j$ as in Figure 2a, are collected in the vector $\mathbf{q}_e = \{q_1, q_2, q_3, q_4, q_5, q_6\}^T$. The dual nodal forces provided by axial forces, shear forces and bending moments at nodes $i$ and $j$, as in Figure 2b, are collected in the vector $\mathbf{Q}_e = \{Q_1, Q_2, Q_3, Q_4, Q_5, Q_6\}^T$. During the inelastic analysis the beam element is subjected to a state determination at pre-established cross sections according to the standard Gauss integration scheme. The chosen Gauss points represent control sections where the plastic constitutive laws are usually integrated according to a incremental approach. The weight associated by the integration procedure to each Gauss point is representative of the length of the beam segment with the reduced stiffness due to the plastic deformations. The correspondence between the weights of the Gauss points and the spatial distribution of the stiffness of the stepped beam in Equation (1) is better explained in what follows.

In Figure 3 an initially homogeneous beam with $n$ Gauss points (control sections or integration points) is depicted. Precisely, in the current study, since the first and last integration points are always chosen coincident with the end sections of the element, the Gauss-Lobatto integration scheme is used. The weights and positions of the Gauss points are indicated as $w_i$ and $x_i^G$, $i = 1, \ldots, n$, respectively.

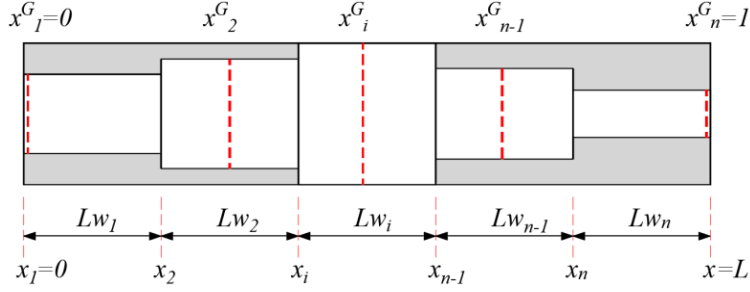

**Figure 3**  Gauss-control integration points and corresponding weights.

In the generic step of the incremental integration procedure in presence of plastic occurrences, the beam can be considered as subdivided into $n$ segments of length $w_i L$ (influence length of the integration point) each characterised by a reduced (tangent) axial stiffness $E_i A_i = E(x_i^G) A(x_i^G)$ and bending stiffness $E_i I_i = E(x_i^G) I(x_i^G)$ evaluated at the relevant Gauss point.

The position $x_i^G$ of the integration points and the length of the reduced stiffness segment $w_i$, according to the Gauss-Lobatto integration procedure, imply abrupt changes of axial and flexural stiffness with discontinuities in axial deformation and the curvature functions at abscissas $x_i = \sum_{j=1}^{i-1} w_j L$, $i = 1, 2, ..., n$.

The abscissas of the flexural stiffness discontinuity occurrences are collected in the vector $\mathbf{x}^{EI} = \{x_1, x_2, ..., x_i, ..., x_n\}^T = \{0, w_1 L, (w_1+w_2)L, ..., (w_1+w_2+...+w_{n-1})L\}^T$. Furthermore, the discontinuity parameters $\beta_{x,i}, \beta_{z,i}$ and $\beta^*_{x,i}, \beta^*_{z,i}$ are collected in the vectors $\boldsymbol{\beta}_x = \{\beta_{x,1}, \beta_{x,2}, ..., \beta_{x,n}\}^T$, $\boldsymbol{\beta}_z = \{\beta_{z,1}, \beta_{z,2}, ..., \beta_{z,n}\}^T$ and $\boldsymbol{\beta}^*_x = \{\beta^*_{x,1}, \beta^*_{x,2}, ..., \beta^*_{x,n}\}^T$, $\boldsymbol{\beta}^*_z = \{\beta^*_{z,1}, \beta^*_{z,2}, ..., \beta^*_{z,n}\}^T$, respectively.

The proposed SDB beam element is characterised by the parameters collected in the vectors $\mathbf{x}^{EI}$ (positions of the axial and flexural stiffness changes) and $\boldsymbol{\beta}^*_x, \boldsymbol{\beta}^*_z$ (intensity of the axial and flexural stiffness changes, respectively) required in Equations (6) and (7) to describe the beam with $n$ stiffness steps.

The shape functions of the axial $u_x(x)$ and transversal $u_z(x)$ displacements for the proposed beam element are determined by imposing the following nodal displacements and rotations:

$$u_x(0) = q_1; \quad u_z(0) = q_2; \quad \varphi(0) = -u'_z(0) = -q_3;$$
$$u_x(L) = q_4; \quad u_z(L) = q_5; \quad \varphi(L) = -u'_z(L) = -q_6 \tag{9}$$

Imposition of the boundary conditions in Equation (9) onto the closed form expression proposed in Equation (6), and its first derivative, leads to the expressions of the axial and transversal deflection in terms of the nodal displacement vector $\mathbf{q}_e$ and the external load function as follows:

$$\mathbf{u}(x; \mathbf{x}^{EI}, \boldsymbol{\beta}^*) = \mathbf{N}(x; \mathbf{x}^{EI}, \boldsymbol{\beta}^*) \cdot \mathbf{q}_e + \mathbf{u}_p(x; \mathbf{x}^{EI}, \boldsymbol{\beta}^*) \tag{10}$$

where the displacement vector $\mathbf{u}^T(x; \mathbf{x}^{EI}, \boldsymbol{\beta}^*) = [u_x(x; \mathbf{x}^{EI}, \boldsymbol{\beta}^*) \quad u_z(x; \mathbf{x}^{EI}, \boldsymbol{\beta}^*)]$ collects the axial and transversal displacement functions dependent on the abscissas and intensity of the stiffness discontinuities collected in the vectors $\mathbf{x}^{EI}, \boldsymbol{\beta}^*$, respectively, as understood by the solution proposed in Equations (6) and (7) for the stepped beam. In Equation (10) $\mathbf{N}(x; \mathbf{x}^{EI}, \boldsymbol{\beta}^*)$ is the shape function matrix defined as follows:

$$\mathbf{N}(x; \mathbf{x}^{EI}, \boldsymbol{\beta}^*) = \begin{bmatrix} N_{x,1}(x; \mathbf{x}^{EI}, \boldsymbol{\beta}^*) & 0 & 0 & N_{x,2}(x; \mathbf{x}^{EI}, \boldsymbol{\beta}^*) & 0 & 0 \\ 0 & N_{z,1}(x; \mathbf{x}^{EI}, \boldsymbol{\beta}^*) & N_{z,2}(x; \mathbf{x}^{EI}, \boldsymbol{\beta}^*) & 0 & N_{z,3}(x; \mathbf{x}^{EI}, \boldsymbol{\beta}^*) & N_{z,4}(x; \mathbf{x}^{EI}, \boldsymbol{\beta}^*) \end{bmatrix} \tag{11}$$

Where $N_{x,k}\left(x;\mathbf{x}^{EI},\boldsymbol{\beta}^*\right)$, $k=1,2$, and $N_{z,j}\left(x;\mathbf{x}^{EI},\boldsymbol{\beta}^*\right)$, $j=1,\ldots,4$, are the axial and transversal displacement shape functions, all dependent on the discontinuity parameters vectors $\mathbf{x}^{EI},\boldsymbol{\beta}^*$, provided explicitly as follows:

$$N_{x,k}(x;\mathbf{x}^{EI},\boldsymbol{\beta}^*) = {}^kA_1 + {}^kA_2 g_2(x), \quad k=1,2$$
$$N_{z,j}(x;\mathbf{x}^{EI},\boldsymbol{\beta}^*) = {}^jC_1 + {}^jC_2 x + {}^jC_3 f_3(x) + {}^jC_4 f_4(x), \quad j=1,\ldots,4 \tag{12}$$

Where

$$\begin{aligned}
&{}^1A_1 = 1; \quad {}^1A_2 = -\frac{1}{g_2(L)}; \quad {}^2A_1 = 0; &&{}^2A_2 = \frac{1}{g_2(L)}\\
&{}^1C_1 = 1; \quad {}^1C_2 = 0; \quad {}^1C_3 = -\frac{f_4^I(L)}{\kappa}; &&{}^1C_4 = \frac{f_3^I(L)}{\kappa};\\
&{}^2C_1 = 0; \quad {}^2C_2 = 1; \quad {}^2C_3 = \frac{-f_4^I(L) + f_4(L)f_2^I(L)}{\kappa}; &&{}^2C_4 = \frac{-f_3(L) + f_3^I(L)}{\kappa};\\
&{}^3C_1 = 0; \quad {}^3C_2 = 0; \quad C_3^3 = \frac{f_4^I(L)}{\kappa}; &&{}^3C_4 = -\frac{f_3^I(L)}{\kappa};\\
&{}^4C_1 = 0; \quad {}^4C_2 = 0; \quad {}^4C_3 = -\frac{f_4(L)}{\kappa}; &&{}^4C_4 = \frac{f_3(L)}{\kappa};
\end{aligned} \tag{13}$$

and

$$\kappa = f_3(L)f_4^I(L) - f_4(L)f_3^I(L) \tag{14}$$

Since the axial and transversal displacement shape functions defined in Equations (12)-(14) depend on the discontinuity parameters, subjected to update during the inelastic analysis, they are named Smart Displacement Shape Functions (SDSFs) in what follows that reminds their ability to follow the state of the beam during the step-by-step non linear analysis.

The last vector in Equation (10) defined as $\mathbf{u}_p^T\left(x;\mathbf{x}^{EI},\boldsymbol{\beta}^*\right) = \left[u_{p_x}\left(x;\mathbf{x}^{EI},\boldsymbol{\beta}^*\right) \quad u_{p_z}\left(x;\mathbf{x}^{EI},\boldsymbol{\beta}^*\right)\right]$, dependent on the vectors $\mathbf{x}^{EI},\boldsymbol{\beta}^*$, provides the additional contributions of the external load distributions $p_x(x), p_z(x)$ to the axial and transversal displacements and it is given as follows:

$$\begin{aligned}
u_{p_x}\left(x;\mathbf{x}^{EI},\boldsymbol{\beta}^*\right) &= -\frac{g_3(L)}{g_2(L)}g_2(x) + g_3(x)\\
u_{p_z}\left(x;\mathbf{x}^{EI},\boldsymbol{\beta}^*\right) &= \frac{f_4(L)f_5^I(L) - f_5(L)f_4^I(L)}{w}f_3(x) + \frac{f_5(L)f_3^I(L) - f_3(L)f_5^I(L)}{w}f_4(x) + f_5(x)
\end{aligned} \tag{15}$$

It has to be noted that the shape function matrix $\mathbf{N}(x;\mathbf{x}^{EI},\boldsymbol{\beta}^*)$, collecting all the SDSFs, and the load vector contribution $\mathbf{u}_p\left(x;\mathbf{x}^{EI},\boldsymbol{\beta}^*\right)$ allow the reconstruction of the element deformed configuration once the nodal displacements are evaluated. Once appropriate shape functions, to be updated during the evolution of non linear events, have been defined it is now possible to introduce the vector of generalised deformation components $\mathbf{d}(x) = \left[\varepsilon_o(x) \quad \chi_y(x)\right]^T$ collecting the axial deformation $\varepsilon_o(x)$ of the beam geometrical axis and the curvature $\chi_y(x)$ of the proposed plane beam element. According to the SDSFs proposed as in Equation (11)-(15), the vector of generalised deformation components $\mathbf{d}(x)$ can be expressed in terms of nodal displacements and distributed external forces, by accounting for the standard Euler-Bernoulli model relationships, as follows:

$$\mathbf{d}(x;\mathbf{x}^{EI},\boldsymbol{\beta}^*) = \mathbf{B}(x;\mathbf{x}^{EI},\boldsymbol{\beta}^*) \cdot \mathbf{q}_e + \tilde{\mathbf{u}}_p\left(x;\mathbf{x}^{EI},\boldsymbol{\beta}^*\right) \tag{16}$$

where the matrix $\mathbf{B}(x;\mathbf{x}^{EI},\boldsymbol{\beta}^*)$, dependent on the non linear state of the element through the discontinuity parameters vectors $\mathbf{x}^{EI},\boldsymbol{\beta}^*$, collects the derivatives of the SDSFs as follows:

$$\mathbf{B}(x;\mathbf{x}^{EI},\boldsymbol{\beta}^*) = \begin{bmatrix} N_{x,1}^{I}(x;\mathbf{x}^{EI},\boldsymbol{\beta}^*) & 0 & 0 & N_{x,2}^{I}(x;\mathbf{x}^{EI},\boldsymbol{\beta}^*) & 0 & 0 \\ 0 & -N_{z,1}^{II}(x;\mathbf{x}^{EI},\boldsymbol{\beta}^*) & -N_{z,2}^{II}(x;\mathbf{x}^{EI},\boldsymbol{\beta}^*) & 0 & -N_{z,3}^{II}(x;\mathbf{x}^{EI},\boldsymbol{\beta}^*) & -N_{z,4}^{II}(x;\mathbf{x}^{EI},\boldsymbol{\beta}^*) \end{bmatrix} \quad (17)$$

It has to be remarked that the functions appearing as elements of the matrix $\mathbf{B}(x;\mathbf{x}^{EI},\boldsymbol{\beta}^*)$ in Equation (17) are discontinuous functions at the cross sections where axial and flexural stiffness undergo abrupt changes in the considered model. Discontinuities are obtained as first and second derivatives of the axial and transversal displacement shape functions, respectively, as formulated in Equation (12) based on the definitions in Equation (8).

The contribution of the external load distributions $p_x(x), p_z(x)$ to the vector of generalised deformation components $\mathbf{d}(x;\mathbf{x}^{EI},\boldsymbol{\beta}^*)$ appearing in Equation (16) is expressed as follows:

$$\tilde{\mathbf{u}}_p\left(x;\mathbf{x}^{EI},\boldsymbol{\beta}^*\right) = \begin{bmatrix} u_{p_x}^{I}\left(x;\mathbf{x}^{EI},\boldsymbol{\beta}^*\right) \\ -u_{p_z}^{II}\left(x;\mathbf{x}^{EI},\boldsymbol{\beta}^*\right) \end{bmatrix} \quad (18)$$

The relationship between the generalised deformation component vector $\mathbf{d}(x;\mathbf{x}^{EI},\boldsymbol{\beta}^*)$ and the normalised internal forces $N(x), M_y(x)$ collected in the vector $\mathbf{D}(x) = \begin{bmatrix} N(x) & M_y(x) \end{bmatrix}^T$ is provided as follows:

$$\mathbf{D}(x;\mathbf{x}^{EI},\boldsymbol{\beta}^*) = \begin{bmatrix} N(x) \\ M_y(x) \end{bmatrix} = \begin{bmatrix} k_{xx}(x) & k_{xz}(x) \\ k_{zx}(x) & k_{zz}(x) \end{bmatrix} \begin{bmatrix} \varepsilon_o(x) \\ \chi_y(x) \end{bmatrix} = \mathbf{k}(x)\,\mathbf{d}(x;\mathbf{x}^{EI},\boldsymbol{\beta}^*) \quad (19)$$

where $\mathbf{k}(x)$ represents the cross section tangent stiffness matrix whose diagonal and off-diagonal elements, $k_{xx}, k_{zz}, k_{xz}, k_{zx}$, rule the direct and cross relations between the internal forces and the deformation components during the inelastic analysis.

The elements of the cross section tangent stiffness matrix $\mathbf{k}(x_i^G)$ evaluated during the non linear step-by-step analysis at the relevant Gauss points $x_i^G, i=1,\ldots,n$, chosen for the integration, will be derived explicitly in the following section by making use of a cross section fibre discretisation approach. However, the elements of the cross section tangent stiffness matrix $\mathbf{k}(x_i^G)$ are related to the discontinuity parameters $\beta_{x,i}, \beta_{z,i}$, introduced in the stepped beam model adopted in Equation (1), according to the following reasoning.

The internal axial force and bending moment increments $dN(x), dM_y(x)$ during the inelastic analysis, in view of Equation (19), depend on the increment of the deformation components as follows:

$$dN(x) = \frac{\partial N(x)}{\partial \varepsilon_o} d\varepsilon_o + \frac{\partial N(x)}{\partial \chi_y} d\chi_y = k_{xx}(x)d\varepsilon_o + k_{xz}(x)d\chi_y$$
$$dM_y(x) = \frac{\partial M_y(x)}{\partial \varepsilon_o} d\varepsilon_o + \frac{\partial M_y(x)}{\partial \chi_y} d\chi_y = k_{zx}(x)d\varepsilon_o + k_{zz}(x)d\chi_y \quad (20)$$

From Equation (20), evaluated at the Gauss integration points at abscissae $x_i^G, i=1,\ldots,n$, the following expressions can be inferred for the tangent axial stiffness $E_i A_i$ and tangent bending stiffness $E_i I_i$ of the $i$-th segment of the multi-stepped beam in Equation (1):

$$E_i A_i = \frac{dN(x_i^G)}{d\varepsilon_o} = k_{xx}(x_i^G) + k_{xz}(x_i^G)\frac{d\chi_y}{d\varepsilon_o}$$
$$E_i I_i = \frac{dM_y(x_i^G)}{d\chi_y} = k_{zx}(x_i^G)\frac{d\varepsilon_o}{d\chi_y} + k_{zz}(x_i^G) \quad (21)$$

The discontinuity parameters $\beta_{x,i}, \beta_{z,i}$ appearing in the SDSFs, defined explicitly in Equations (11)-(13), can be obtained by Equation (21), accounting for their relations $\beta_{x,i} = \dfrac{E_o A_o - E_i A_i}{E_o A_o}$, $\beta_{z,i} = \dfrac{E_o I_o - E_i I_i}{E_o I_o}$ with the axial stiffness $E_i A_i$, $i = 1,\ldots,n$, and the flexural stiffness $E_i I_i$, $i = 1,\ldots,n$, respectively, as follows:

$$\begin{aligned}\beta_{x,i} &= 1 - \frac{1}{E_o A_o}\left[k_{xx}(x_i^G) + k_{xz}(x_i^G)\frac{d\chi_y}{d\varepsilon_o}\right] \\ \beta_{z,i} &= 1 - \frac{1}{E_o I_o}\left[k_{zx}(x_i^G)\frac{d\varepsilon_o}{d\chi_y} + k_{zz}(x_i^G)\right]\end{aligned} \qquad (22)$$

The discontinuity parameters $\beta_{x,i}, \beta_{z,i}$ can be updated during the inelastic analysis by making use of Equation (22) in terms of the elements $k_{xx}, k_{zz}, k_{xz}, k_{zx}$ of the cross section tangent stiffness matrix $\mathbf{k}(x_i^G)$ evaluated at the Gauss integration points, the latter being evaluated according to cross section fibre discretisation approach as better specified in the next section.

Being the shape functions related to the Euler-Bernoulli beam model, the internal shear force $T_z(x)$ has not been included in the internal force vector $\mathbf{D}(x)$ since it is not directly evaluated by the integration of the non linear constitutive equations but rather evaluated by equilibrium conditions at the end of each iteration.

## THE FE DISCRETISATION BY MEANS OF A FIBRE APPROACH FOR R/C CROSS SECTIONS: FIBER SMART DISPLACEMENT BASED (FSDB) APPROACH

The formulation of the SDB beam element presented in section 3 is mainly oriented towards the non linear analysis of frame structure with r/c cross sections characterised by area partialization, due to the low concrete tensile strength, and plastic deformation of the steel bars. For these reasons, rather than adopting a sectional approach, a sectional fibre approach is exploited and suitably adapted to the proposed SDSFs. According to the fibre approach, each Gauss cross section is discretised into small areas, denoted as fibres, that may be thought as stripes or squares for 2D and 3D problems, respectively. Each fibre is characterised by a different nonlinear uniaxial law in accordance to the material constitutive behaviour. Parallel integration of the non linear constitutive laws at each fibre in the step-by-step analysis provides the cross section internal forces in the corrector phase.

The adopted fibre approach coupled with the use of the SDSFs allows a suitable modelling of the axial-bending interaction for the case of r/c cross sections.

A r/c cross section area is hence discretised into $n_c$ concrete fibres, as depicted in Figure 4 in the form of stripes and $n_b$ steel fibres. Each fibre is characterised by an area $A_c$ (concrete) or $A_s$ (steel), where $c = 1,\ldots,n_c$ and $s = 1,\ldots,n_b$. The generic $f$-th fibre is characterised by an area $A_f$ ($f = 1,\ldots,n_f = n_c + n_b$) and a non linear uniaxial normal stress-strain constitutive behaviour since shear deformations and the consequent interaction with axial deformations is not taken into account. By assuming the principle of planar section conservation the axial strain $\varepsilon_x(x)$ of each fibre is written as:

$$\varepsilon_x(x; z_f) = \varepsilon_o(x) + \chi_y(x) z_f = \begin{bmatrix} 1 & z_f \end{bmatrix} \begin{bmatrix} \varepsilon_o(x) \\ \chi_y(x) \end{bmatrix} = \boldsymbol{\alpha}(z_f) \cdot \mathbf{d}(x) \qquad (23)$$

Where the row vector $\boldsymbol{\alpha}(z_f) = \begin{bmatrix} 1 & z_f \end{bmatrix}$, dependent on the distance $z_f$ of the $f$-th fibre from the beam axis, has been introduced. The axial deformation of the $f$-th fibre of the cross section can be expressed in terms of nodal displacements $\mathbf{q}_e$ for the proposed SDB beam element by replacing Equation (16), purged of the external load contribution $\tilde{\mathbf{u}}_p\left(x; \mathbf{x}^{EI}, \boldsymbol{\beta}^*\right)$, into Equation (23) as follows:

$$\varepsilon_x(x; z_f) = \boldsymbol{\alpha}(z_f) \cdot \mathbf{B}(x; \mathbf{x}^{EI}, \boldsymbol{\beta}^*) \cdot \mathbf{q}_e \qquad (24)$$

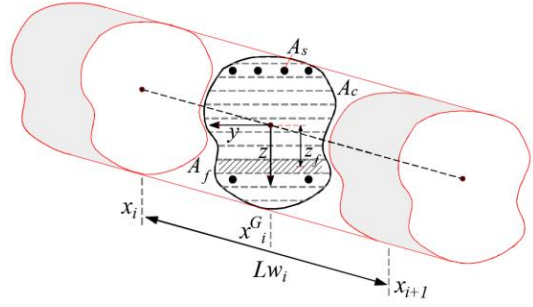

**Figure 4**     Fibre discretization of a r/c cross section according to a 2D formulation.

The non linear uni-axial constitutive relation between the axial strain $\varepsilon_x(x;z_f)$ and the normal stress $\sigma_x(x;z_f)$ can written as $\sigma_x(x;z_f) = E_T(x;z_f)\varepsilon_x(x;z_f)$ in terms of the tangent modulus $E_T(x;z_f)$ of the *f*-th fibre at cross section with abscissa $x$ to be evaluated according to the uni-axial constitutive model chosen for the applications.

**The element stiffness matrix**

The principle of virtual displacements for the SDB beam presented in section 3, where the above described cross section fibre discretisation has been introduced, writes:

$$\mathbf{Q}_e^T \cdot \delta\mathbf{q}_e = \int_0^1 \sum_{f=1}^{n_f} \sigma_x(x;z_f) A_f \delta\varepsilon_x(x;z_f)\, dx \qquad \forall \delta\mathbf{q}_e, \delta\varepsilon_x(x) \tag{25}$$

where $\delta$ indicates virtual quantities. Substitution of the relation expressed by Equation (24), together with the relevant normal stress-axial strain relationship, into Equation (25) leads to the following expression of the principle of virtual displacements in terms of nodal displacements $\mathbf{q}_e$:

$$\mathbf{Q}_e^T \cdot \delta\mathbf{q}_e = \mathbf{q}_e^T \int_0^1 \mathbf{B}^T(x;\mathbf{x}^{EI},\boldsymbol{\beta}^*) \sum_{f=1}^{n_f} \boldsymbol{\alpha}^T(z_f) E_T(x;z_f) A_f \boldsymbol{\alpha}(z_f)\, \mathbf{B}(x;\mathbf{x}^{EI},\boldsymbol{\beta}^*) dx\, \delta\mathbf{q}_e(x;z_f) \qquad \forall \delta\mathbf{q}_e \tag{26}$$

Equation (26) implies the following relationship between the nodal force $\mathbf{Q}_e$ and nodal displacement $\mathbf{q}_e$ vectors:

$$\mathbf{Q}_e = \mathbf{K}_e(\mathbf{x}^{EI},\boldsymbol{\beta}^*) \cdot \mathbf{q}_e \tag{27}$$

Where $\mathbf{K}_e(\mathbf{x}^{EI},\boldsymbol{\beta}^*)$ is the element stiffness matrix, dependent on the current state of the element by means of the plastic intensity parameter vector $\boldsymbol{\beta}^*$ and the plastic segment extension vector $\mathbf{x}^{EI}$, defined as follows:

$$\mathbf{K}_e(\mathbf{x}^{EI},\boldsymbol{\beta}^*) = \int_0^1 \mathbf{B}^T(x;\mathbf{x}^{EI},\boldsymbol{\beta}^*)\mathbf{k}(x)\mathbf{B}(x;\mathbf{x}^{EI},\boldsymbol{\beta}^*)\, dx \tag{28}$$

The inner matrix $\mathbf{k}(x)$ appearing in the integral in Equation (28) represents the cross section stiffness matrix obtained by the adopted fibre discretisation as follows:

$$\mathbf{k}(x) = \sum_{f=1}^{n_f} \boldsymbol{\alpha}^T(z_f) E_T(x;z_f) A_f \boldsymbol{\alpha}(z_f) = \begin{bmatrix} \sum_{f=1}^{n_f} E_T(x;z_f)A_f & \sum_{f=1}^{n_f} E_T(x;z_f)A_f z_f \\ \sum_{f=1}^{n_f} E_T(x;z_f)A_f z_f & \sum_{f=1}^{n_f} E_T(x;z_f)A_f z_f^2 \end{bmatrix} \tag{29}$$

The cross section stiffness matrix $\mathbf{k}(x)$ in Equation (29) is the fibre counterpart of the expression reported in Equation (19). The components of the cross section stiffness matrix $\mathbf{k}(x)$ provided by Equation (29) are related to the discontinuity parameters collected in the vector $\boldsymbol{\beta}$ and the physical and geometrical properties of fibre discretised cross section as in Equation (22). Precisely, in view of Equations (28) and (22) the following relationships hold:

$$\beta_{x,i} = 1 - \frac{1}{E_o A_o} \left[ \sum_{f=1}^{n_f} E_T(x_i^G; z_f) A_f + \sum_{f=1}^{n_f} E_T(x_i^G; z_f) A_f z_f \frac{d\chi_y}{d\varepsilon_o} \right]$$

$$\beta_{z,i} = 1 - \frac{1}{E_o I_o} \left[ \sum_{f=1}^{n_f} E_T(x_i^G; z_f) A_f z_f \frac{d\varepsilon_o}{d\chi_y} + \sum_{f=1}^{n_f} E_T(x_i^G; z_f) A_f z_f^2 \right]$$

(30)

Equation (30) is used to retrieve the values of the discontinuity parameters collected in the vector $\boldsymbol{\beta}$ during the non linear analysis once the integration of the non linear constitutive equations is performed for each fibre of the cross sections. The expression reported in Equation (28) shows clearly how the element stiffness matrix $\mathbf{K}_e$, differently from the classical displacement based approach commonly adopted in the literature, depends on the variation of the shape functions which are updated in accordance to the discontinuity parameter vector $\boldsymbol{\beta}^*$.

According to the Gauss integration scheme the element stiffness matrix in Equation (28) can be evaluated as follows:

$$\mathbf{K}_e(\mathbf{x}^{EI}, \boldsymbol{\beta}^*) \approx L \sum_{r=1}^{n} \mathbf{B}^T(x_r^G; \mathbf{x}^{EI}, \boldsymbol{\beta}^*) \mathbf{k}(x_r^G) \mathbf{B}(x_r^G; \mathbf{x}^{EI}, \boldsymbol{\beta}^*) w_r \qquad (31)$$

Where the matrix $\mathbf{B}(x_r^G; \mathbf{x}^{EI}, \boldsymbol{\beta}^*)$ collecting the derivatives of the displacement shape functions formulated explicitly in Equations (12)-(14) are evaluated at the Gauss integration cross sections and updated at each iteration as plastic deformations occur. The inner matrix $\mathbf{k}(x_i^G)$ in the product appearing in Equation (31) represents the tangent stiffness matrix of the Gauss cross sections to be evaluated by means of integration of the fibre constitutive equations across the cross section in accordance with Equation (29).

**The resisting nodal forces and the external nodal forces**

Besides the formulation of the SDB beam element stiffness matrix, FE analysis based on a displacement approach requires the enforcement of the structural nodal equilibrium conditions. To this purpose the element nodal resisting forces $\mathbf{Q}_e$ can be determined by replacing Equation (24) in the expression of the principle of virtual displacements as formulated in Equation (25):

$$\mathbf{Q}_e^T \cdot \delta \mathbf{q}_e = \int_0^1 \sum_{j=1}^{n_f} \sigma_x(x; z_f) A_f \, \boldsymbol{\alpha}(z_f) \cdot \mathbf{B}(x; \mathbf{x}^{EI}, \boldsymbol{\beta}^*) dx \, \delta \mathbf{q}_e \qquad \forall \delta \mathbf{q}_e$$

(32)

Equation (32) implies the following definition for the element nodal resisting forces in terms of internal stress distribution:

$$\mathbf{Q}_e(\mathbf{x}^{EI}, \boldsymbol{\beta}^*) = \int_0^1 \mathbf{B}^T(x; \mathbf{x}^{EI}, \boldsymbol{\beta}^*) \sum_{j=1}^{n_f} \sigma_x(x; z_f) A_f \, \boldsymbol{\alpha}^T(z_f) \, dx \qquad (33)$$

Where the summation term appearing in the integral represents the vector of internal forces $\mathbf{D}(x) = \begin{bmatrix} N(x) & M_y(x) \end{bmatrix}^T$ explicitly written as follows:

$$\mathbf{D}(x) = \begin{bmatrix} N(x) \\ M_y(x) \end{bmatrix} = \sum_{j=1}^{n_f} \sigma_x(x, z_f) A_f \, \boldsymbol{\alpha}^T(z_f) \qquad (34)$$

The element nodal resisting forces $\mathbf{Q}_e$, provided by Equation (33), can be numerically evaluated according to the Gauss integration scheme as follows:

$$\mathbf{Q}_e(\mathbf{x}^{EI}, \boldsymbol{\beta}^*) \approx L \sum_{r=1}^{n} \mathbf{B}^T(x_r^G, \mathbf{x}^{EI}, \boldsymbol{\beta}^*) \mathbf{D}(x_r^G) w_r \qquad (35)$$

The internal force vector $\mathbf{D}(x_r^G)$ is computed at the Gauss cross sections by means of Equation (34) where the normal stresses $\sigma_x(x_r^G; z_f)$ are evaluated in the corrector phase through integration of the fibre constitutive equations. On the other hand, the nodal forces equivalent to the along axis external forces, to be balanced by the internal forces determined by means of Equation (35), are given as:

$$\mathbf{P}_e(\mathbf{x}^{EI}, \boldsymbol{\beta}^*) = \int_0^1 \mathbf{N}^T(x; \mathbf{x}^{EI}, \boldsymbol{\beta}^*) p(x)\, dx \tag{36}$$

The element nodal external forces $\mathbf{P}_e$, provided by Equation (36), can be numerically evaluated according to the Gauss integration scheme as follows:

$$\mathbf{P}_e(\mathbf{x}^{EI}, \boldsymbol{\beta}^*) \approx L \sum_{r=1}^{N} \mathbf{N}^T(x_r^G, \mathbf{x}^{EI}, \boldsymbol{\beta}^*) p(x_r^G)\, w_r \tag{37}$$

## AXIALLY EQUILIBRATED SDB BEAM ELEMENT

One of the problems encountered in the formulation of the classical DB beam element is due to the assumption of the linear axial shape functions implying a constant axial force distribution during the analysis which does not reflect the variation of the internal axial force evaluated at the Gauss cross sections during the integration of the non linear constitutive laws. As a result, axial equilibrium is not strictly enforced along the beam axis and it is rather verified in a weak form.

On the other hand, the FB approach, based on adoption of both constant and linear axial force shape functions does not suffer such approximation and leads to exact solutions regardless of the non linear constitutive behaviour.

The lack of strong form axial equilibrium of classical DB elements, verified in average sense only, represents a strong limitation with respect the FB approach which can be overcome by adopting a mesh refinement. The latter issue, clearly discussed in [Calabrese et al. 2010], implies a low performance of the DB element and requires a dense mesh to reach an accuracy comparable to the FB approach against experimental results. On the other hand, the concept of DB beam element strictly satisfying axial equilibrium along the beam axis was originally introduced by Izzudin et al. [Izzudin et al. 1994] for non-linear elastic problems proposing the use of higher-order quartic shape functions. In the latter paper it is clearly remarked how the use of linear axial displacement shape functions may lead to a low accurate DB element with an over stiffness. On the other hand, an interesting and appealing approach to achieve strong axially equilibrium for DB beam elements was recently proposed by Tarquini et al. [Tarquini et al. 2017] in the context of distributed plasticity. In the latter work the discrepancy of different axial force values at the Gauss points (due to the integration of plastic constitutive laws) with the constant axial force (due to the linear axial displacement shape functions assumption) is resolved by means of a suitable correction of the sectional axial strains. Precisely, an internal iterative procedure to correct the axial strains at each Gauss point is triggered until the same value of the axial force along the element is ensured. The correction of the axial strain is computed at each iteration by solving a linear system of equations correspondent to the number of Gauss points.

Given the above premises, the SDSFs proposed in section 3 can be shown to be efficaciously employed to formulate an axially equilibrated SDB beam element (SDB/ae) in a strong form.

The idea behind the formulation of a SDB/ae beam element relies on the evaluation of a point load distributions, generated by the axial force difference between each two successive Gauss points, and the imposition of an axial deformation distribution increment obtained by using the SDSFs formulated in Equations (10)-(15). In practice, at the generic iteration of the Newton-Raphson procedure, when the plastic constitutive laws are independently integrated at each Gauss point (sectional force updating) the difference of axial force $N(x_i^G) - N(x_{i-1}^G)$, $i = 1,\ldots,n$, between successive Gauss points is re-imposed over the element under the form of the following fictitious axial load distribution of concentrated loads:

$$\tilde{p}_x(x) = \sum_{i=1}^{n-1} \left[ N(x_{i+1}^G) - N(x_i^G) \right] \delta(x - x_{i+1}) \tag{38}$$

where $\delta(x - x_i)$, the distributional derivative of the Heaviside (unit step) generalised function $U(x - x_i)$ well known as Dirac's delta distribution, is adopted in Equation (38) to model a sequence of axial point loads concentrated at cross sections $x_i$, $i = 1,\ldots,n-1$ (Figure 5).

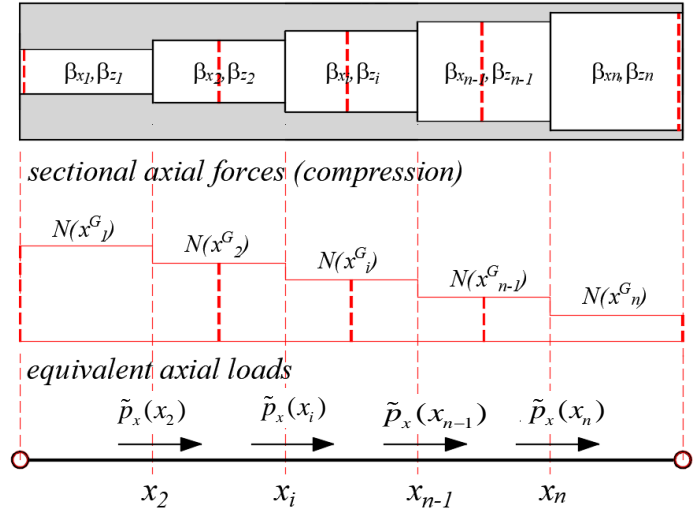

**Figure 5**  Longitudinal forces distribution equivalent to the internal axial unbalance.

The axial load distribution $\tilde{p}_x(x)$, as defined in Equation (38), is responsible for the onset of the axial displacement increment field $\Delta u_{\tilde{p}_x}\left(x; x^{EI}, \beta^*\right)$ straightforwardly provided by the first expression in Equation (15) as follows:

$$\Delta u_{\tilde{p}_x}\left(x; x^{EI}, \beta^*\right) = -\frac{g_3(L)}{g_2(L)} g_2(x) + g_3(x) \tag{39}$$

With, in view of the definitions in Equation (7),

$$\begin{aligned}
g_2(x) &= -x - \sum_{i=1}^{n} \beta_{x,i}^* \left(x - x_i\right) U(x - x_i) \\
g_3(x) &= -\frac{\tilde{p}_x^{[2]}(x)}{E_o A_o} - \sum_{i=1}^{n} \frac{\beta_{x,i}^*}{E_o A_o} \left[\tilde{p}_x^{[2]}(x) - \tilde{p}_x^{[2]}(x_i)\right] U(x - x_i)
\end{aligned} \tag{40}$$

The use of additional internal force, for improving the element response, leads to an along beam distribution of fictitious concentrated axial loads that could appear physically inconsistent. However this strategy is justified by the need to relax the axial locking related to the use of linear axial shape functions along the beam and re-establish the axial equilibrium along the beam. This correction appear particularly important because wrong values of axial force at the control points lead to erroneous evaluations of the neutral axis positions and the corresponding limit bending moment.

It has to be noted that, in accordance to the idea behind the SDSFs, the dependence of $\Delta u_{\tilde{p}_x}\left(x; x^{EI}, \beta^*\right)$ on the vectors $x^{EI}, \beta^*$, representing the state of the beam undergoing distributed plasticity during the Newton-Raphson iterative strategy, has been evidenced. The axial displacement increment field $\Delta u_{\tilde{p}_x}\left(x; x^{EI}, \beta^*\right)$, due to the axial point distribution $\tilde{p}_x(x)$ in Equation (38), gives rise to a fictitious axial deformation increment $\Delta \varepsilon_{\tilde{p}_x}(x; x^{EI}, \beta^*) = \frac{d}{dx} \Delta u_{\tilde{p}_x}(x; x^{EI}, \beta^*)$ responsible of an inner iterative correction of the trial axial deformation until a constant axial force is obtained over the element delivering an axially equilibrated element in a strong form.

In other words, within the standard predictor-corrector Newton-Raphson procedure, after the element state determination an internal iterative procedure is triggered by the superimposition of the axial deformation increment $\Delta \varepsilon_{\tilde{p}_x}(x; x^{EI}, \beta^*)$ due to the lack of axial force equilibrium in strong form. When the increment $\Delta \varepsilon_{\tilde{p}_x}(x; x^{EI}, \beta^*)$ is applied a new element state determination is performed to determine the updated axial force unbalance. The inner iterations are stopped when the difference of axial forces $N(x_i^G) - N(x_{i-1}^G)$, $i = 1, \ldots, n$, is null (within the accepted tolerance), delivering a strongly axially equilibrated element.

The above defined axial deformation increment distribution, imposed iteratively to generate an axially equilibrated beam element, has been obtained in closed form consistently with the distributed plasticity model adopted in the formulation and with the nodal displacement given in the current iteration.

The presented procedure to obtain an axially equilibrated element differs from that proposed in [Tarquini et al. 2017] since a two condition algebraic system is therein solved at each iteration to obtain the axial corrections while, following the closed form formulation of the beam element presented in section 2 and the relevant SDSFs, the latter are provided explicitly. Another source of difference between the two procedures is related to the smart (i.e. adaptive) nature of the shape functions that are tailored to the state of the beam that is related to axial and flexural inelastic response.

## MODEL VALIDATION

The proposed model is validated considering simple benchmarks represented by cantilever beams subjected to a constant axial compression force and variable bending moment related to an increasing concentrated force applied at the free end of the cantilever. All the considered benchmarks have been experimentally and/or numerically tested in literature under monotonic or cyclic pseudo-static actions. In the following the results obtained by means of the proposed FSDB beam model are also compared to the standard DB model, the DB axial equilibrated (DB/ae) model [Tarquini et al. 2017] and the Force Based (FB) element, this latter implemented in the software OpenSees [McKenna 2011]. The analyses employing the FSDB, DB and DB/ae model are performed through a direct implementation in an original software developed within the general object-oriented programming language Visual C# [Microsoft Visual Studio 2013] and denominate "*SMART-FRAME*" in which the proposed model has been implemented. The analyses based on the FB model are performed in OpenSees [McKenna 2011] employing the "Concrete02" and "Steel02" uniaxial materials available in OPENSEES. The simulations are performed using ten Gauss-Lobatto Integration Points (IPs).

### Benchmark 1

The benchmark considered in this section is a reinforced concrete cantilever beam already studied in [Tarquini et al. 2017] by means of standard and axial equilibrated displacement elements, and previously by McKenna [McKenna 2011] considering the forced based element model, implemented in the software OpenSees. The reinforced concrete beam has a length $L = 300 cm$, rectangular cross section $30 \times 40 cm$ and $20 mm$ of cover concrete. Twelve $\phi 16$ mm reinforcing steel bars are symmetrically disposed along the section, as reported in Figure 60a. The analyses are conducted applying an increasing shear force $F$, at the free end of the beam, after the application of a constant axial force, $N = 75 kN$, corresponding only to 1.25% of the axial-compression capacity of the section.

In the numerical simulations the modified Kent and Park constitutive law [Kent and Park 1971], according to Yassin model [Yassin 2011], is adopted in compression while a linear softening behaviour is considered when the tensile strength is reached. The compression strength of the confined concrete is $f_{cc} = 42 \text{Mpa}$, while the strength of the unconfined concrete is $f_c = 37 \text{Mpa}$. The concrete strain at the peak-stress are respectively $\varepsilon_c = 0.24\%$ and $\varepsilon_{cc} = 0.28\%$ for unconfined and confined concrete. For both the concrete material an initial Young module $E_c = 30 GPa$ of the Kent and Park model [Kent and Park 1971] is considered. The tensile strengths of the concrete materials are assumed equal to 10% of the corresponding compression strengths and the softening slope is equal to $2/3 E_c$. The steel bars are modelled according to the Menegotto and Pinto constitutive law [Menegotto and Pinto 1973] with Young modulus $E_S = 200 GPa$, $f_y = 480 MPa$, hardening $0.5\%$, initial value of the curvature parameter 15 and curvature degradation parameters respectively 0.925 and 0.15. All the analyses are performed by discretising the cross section by means of 40 fibres.

Figure 6b reports the results obtained discretising the cantilever beam with single FB, DB and FSDB beam element and ten Gauss points consistently to a Gauss Legendre integration. The capacity curves in terms of external force $F$ versus the deflection of the free end of the beam $u_z(L)$ are reported.

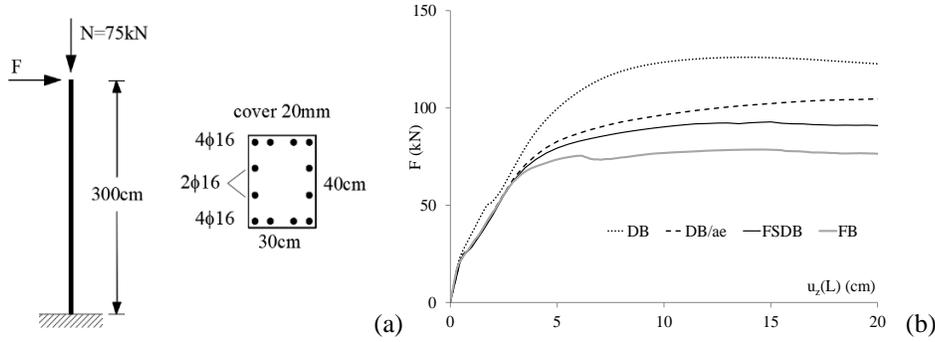

**Figure 6**  Cantilever geometrical layout (a) and comparison of the force-deflection curves (b).

The maximum lateral force registered by the FB model is 78.6kN. The standard DB model overestimates the beam lateral strength of 60.1% (126.0kN), while the error decreases to 33.2% and 18.1% by using the DB/ae model and the FSDB model, respectively, so providing a more reliable prediction of the load-caring capacity of the beam with the use of a single frame element.

Figures 7 show the curvature $\chi_y$, the axial strain $\varepsilon_o$, the $\beta_z$ and $\beta_x$ parameter distributions, as a function of the normalised beam axis, for a single SFSDB model with 10 IPs, corresponding to three different levels of normalized drift $\delta$ (normalised transversal displacement at the free end). Figure 7a reports the multi-linear distribution of the curvature related to the distribution of plasticity along the beam. The axial strain variability is reported in Figure 7b this is the consequence of the additional axial concentrated forces and the variation of axial stiffness along the beam length. The influence of the $\beta_z$ parameters on the response of the beam is visible by the curvature distribution, a multilinear trend can be observed due to the ability of the smart element to adapt the flexural stiffness distribution to the state of the beam. Similarly the $\beta_x$ parameters produce a stepped non linear response in terms of axial strains, being the values related to the Gauss points and the corresponding beam portions.

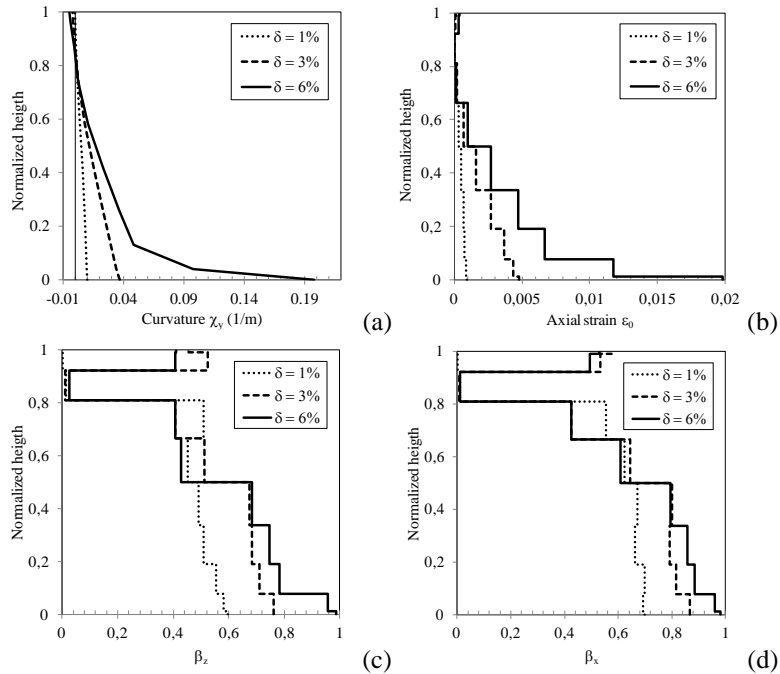

**Figure 7**  FSDB model: Curvature (a) and axial strain (b) distributions, $\beta_x$ (c) and $\beta_z$ (d) parameters.

Figure 8 shows sensitivity of the FSDB model to the number of the IPs when a single element is employed (0a) and the number of elements adopted for the beam discretisation (Figure 8b). In the same Figure, the lines representative of the FB and standard DB responses are also reported. A not significant influence on the response is observed by increasing the IPs from 5 to 20, while by increasing the number of elements the FSDB response converges to the FB curve with 10 IPs.

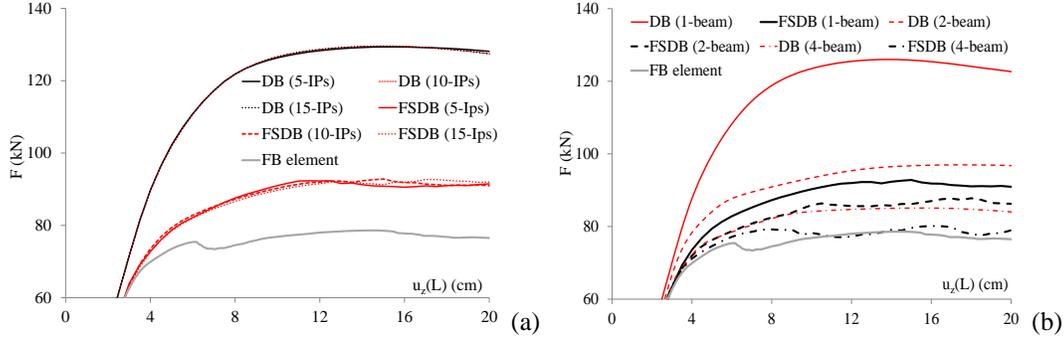

**Figure 8**  Influence of the number of elements with 10 IPs (a) and the number of the IPs (b).

In Figure 9 the comparison of the cyclic responses of the beam obtained by means of the standard DB model, the FSDB model and the FB model is reported using a single element. The beam is subjected to an imposed cyclical displacement at the free end $u_z(L)$ whose amplitude is increased of 30mm at each loading branch, up to the maximum displacement of 210mm. The reaction ($F$) at the restrained end is plotted in the graph. The peak transversal force of the system registered by the FB model was 73.91kN and -72.67kN in the positive and negative loading phase respectively. The corresponding peak-load values obtained by the DB model were 120.60kN and -119.14kN with an error of 63.18% with respect to the referred FB model. Furthermore the DB model overestimates the post-peak softening behaviour if compared to the other models. The proposed FSDB model provide peaks equal to 83.2kN and -80.45kN respectively associated to a maximum error of 12.6%, compared to the FB model. In terms of hysteretic behaviour, the FSDB model is able to reproduce the pinching effect due to the closing of the concrete tensile cracking, coherently to the FB model and completely neglected by the DB model. However, the figure shows that the FSDB model leads to a weak overestimation of the hysteretic area related to the reloading phases however, these differences do not lead to significant consequences and the system envelope cyclical response is well predicted.

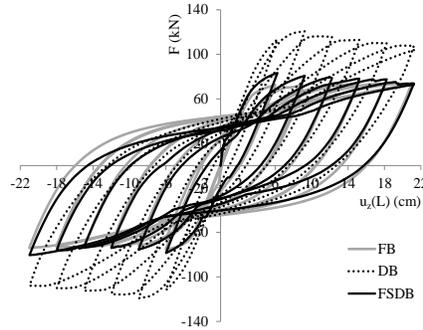

**Figure 9**  Transversal force versus lateral deflection $u_z(L)$ for a cyclic analysis.

**Benchmark 2: the Low and Moehle beam**
The second benchmark is a reinforced concrete cantilever beam experimentally tested by Low and Moehle [Low and Moehle 1987] and numerically investigated by Spacone and Filippou [Spacone et al. 1996b] using their proposed FB model. During the experimental research different prototypes, subjected to uni-axial and biaxial bending moments under different levels of compression axial load, were studied. Here the results regarding the *specimen 1* are numerically simulated. The beam has length $L = 51.44$cm and a rectangular cross section $12.7 \times 16.5$cm with ten uniformly distributed reinforcing steel bars, Figure 10a. The beam is subjected to a uniaxial bending moment, along the direction of minor inertia (axis z in Figure 10a), and a constant axial load $N = 44.48$kN corresponding approximately to 5% of the ultimate compression-load capacity of the concrete. The mechanical parameters are fixed according to [Spacone et al. 1996b]. The same constitutive laws of benchmark 1 are employed in the analysis with zero tensile concrete strength. The confined concrete is characterised by an ultimate compression stress $f_{cc} = 42$MPa, residual stress $21$MPa, peak-stress and ultimate strain $\varepsilon_{cc} = 0.23\%$, $\varepsilon_{ccu} = 23.3\%$. The unconfined concrete is characterised by the peak and ultimate compression stress $f_c = 37$MPa, $17$MPa, and strain $\varepsilon_c = 0.20\%$, $\varepsilon_{cu} = 1.19\%$. The steel is characterised by $f_y = 447$MPa, $E_s = 200$GPa and hardening factor 0.0067.

In Figure 10a is shown the test geometry of the beam, the transversal section and the load conditions, while in Figure 10b is reported the comparison between the numerical response obtained by the proposed and the experimental results expressed in terms of deflection of the loaded point versus the applied load. A single element and two element mesh are employed and 10 IPs for each element. A satisfactory agreement can be observed both in terms of envelope curve and hysteretic behaviour. The numerical responses are characterised by a higher initial stiffness since the model does not take into account the bond-slip of the reinforcing steel bars. However a good prediction of the global response is obtained.

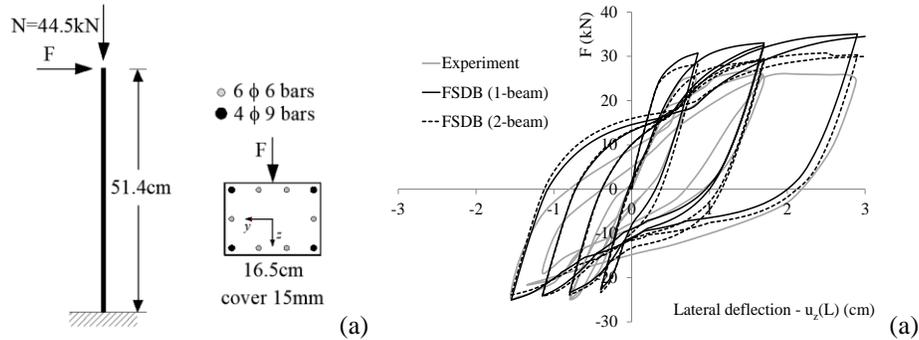

**Figure 10** Low and Moehle beam: test layout (a); numerical vs experimental response (b).

In Figure 11a the comparison of the cyclic responses of the beam obtained by means the standard DB model, the FSDB model and the FB model is reported. A cyclical displacement is imposed at the free end of the beam with constant step of 30mm, up to the maximum displacement of 21mm. It can be observed that the FSDB provides a satisfactory prediction in terms of envelope curve, while the standard DB model leads to a significant error in terms of over-strength. A certain overestimation of the positive reloading branch is also observed for the FSDB model. Figure 11b shows the axial deformations of the baricentre of the free end of the beam. It can be observed that the proposed model provides a good prediction of the elongation of the beam due to the concrete partialization. On the contrary, the standard DB model overestimates these deformations. This latter aspect could be particularly important for the simulation of framed structures.

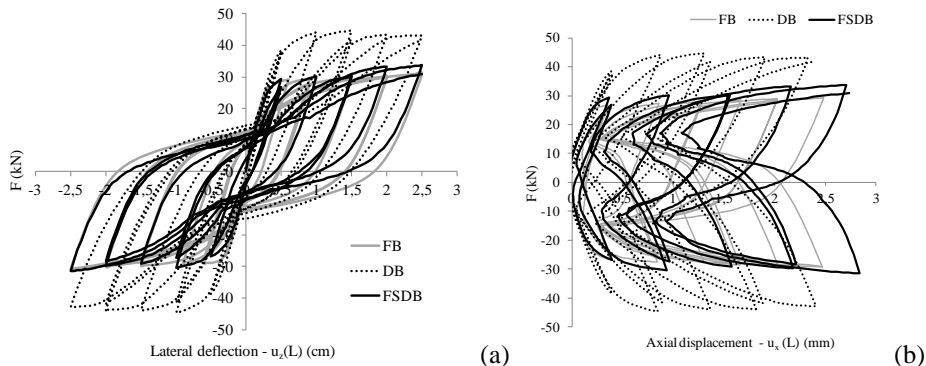

**Figure 11** Low and Moehle beam: test layout (a); External force F versus (b) $u_z(L)$ ; (c) $u_x(L)$.

**Benchmark 3: the Kent beam**
The present section aims at evaluating the performance of the proposed model on simulating the bending moment-axial force interaction for different levels of the axial compression load. To this aim, the cantilever beam experimentally tested by [Kent 1969] and numerically investigated in [Spacone et al. 1996b] is considered. The beam, depicted in Figure 12a, has length $L = 225$cm and $12.3 \times 20.3$cm rectangular cross section. Only Four reinforcing steel bars are disposed at the vertexes of the section with 25 mm of top and bottom concrete cover and 20 mm of lateral cover. The same constitutive laws used for the benchmark 1 are considered. The concrete ultimate compression stress and the corresponding strain are $49.7 MPa$ and $0.27\%$ for confined and unconfined concrete respectively and zero-tensile strength. The ultimate strain values are assumed to be $0.30\%$ and $3.00\%$ for unconfined and confined concrete. The steel is characterised by $f_y = 336 MPa$, $E_s = 200 GPa$ and hardening factor 0.0042. Figure 12b shows the bending moment - curvature cyclic behaviour of the restrained section of the beam obtained by a structural analysis of the cantilever beam employing the FSDB model and imposing at the free end two cycles with amplitude 45mm and 60mm respectively, in absence of compression load. This response is compared to the numerical moment-curvature provided by [Spacone et al. 1996] and [Kent 1969].

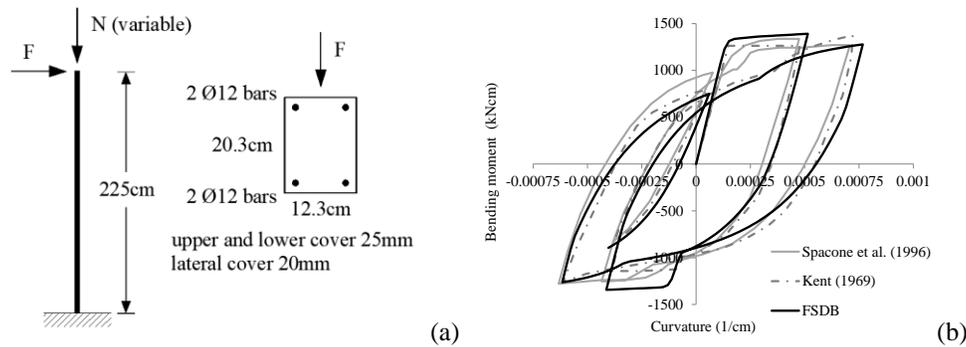

**Figure 12. Kent's beam [A]: (a) geometrical layout; (b) bending moment – curvature cyclic behaviour.**

The ultimate capacity loads obtained by a single DB and FSDB element for different levels of the axial load $N = 0, 75, 150, 230, 300, 375\ kN$ are summarised in Table 1. In the same table the difference ($\delta$) of the DB and FSDB models, compared to the FB predictions, are evaluated. In Figure 13 the capacity curves in terms of external force $F$ versus the lateral deflection of the free end of the column $u_z(L)$ are compared considering a single element and 2 elements mesh discretization.

**Table 1    Peak-load registered by the different models and the error to the referred FB model**

| | Model | Axial load – N (kN) | | | | | |
|---|---|---|---|---|---|---|---|
| | | 0 | 75 | 150 | 225 | 300 | 375 |
| $F_u$ (kN) | DB | 12.15 | 16.95 | 19.38 | 20.71 | 21.37 | 21.69 |
| | FSDB | 8.28 | 12.57 | 14.84 | 17.28 | 18.11 | 18.42 |
| | FB | 6.97 | 9.60 | 12.11 | 13.91 | 15.19 | 15.74 |
| $\delta$ (%) | DB | 74.30 | 76.57 | 60.02 | 48.88 | 40.69 | 37.80 |
| | FSDB | 18.68 | 31.01 | 22.53 | 24.21 | 19.26 | 17.03 |

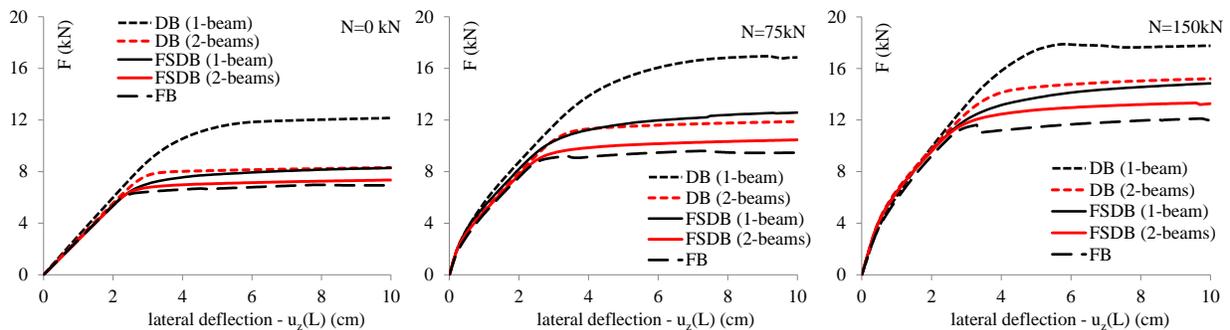

**Figure 13    External force versus the lateral beam deflection for different levels of the axial load.**

Figures 14 report the cyclic behaviour of the beam in terms of shear force $F$ versus the lateral deflection $u_z(L)$ for three different levels of the axial load $N = 0, 75, 150\ kN$. The analyses are performed imposing a transversal displacement of the free end of the column. Five complete cycles are performed increasing the displacement amplitude by 15mm at each cycle and applying a maximum displacement of 75mm. The FSDB results are compared to the ones obtained by the DB and FB model. The comparison shows a satisfactory agreement between the FSDB and the FB solution not only in terms of envelope curve but also in terms of hysteretic cycles. The comparison clearly shows the better performance of the FSDB model compared to the standard DB element whose accuracy worsens for low values of axial load.

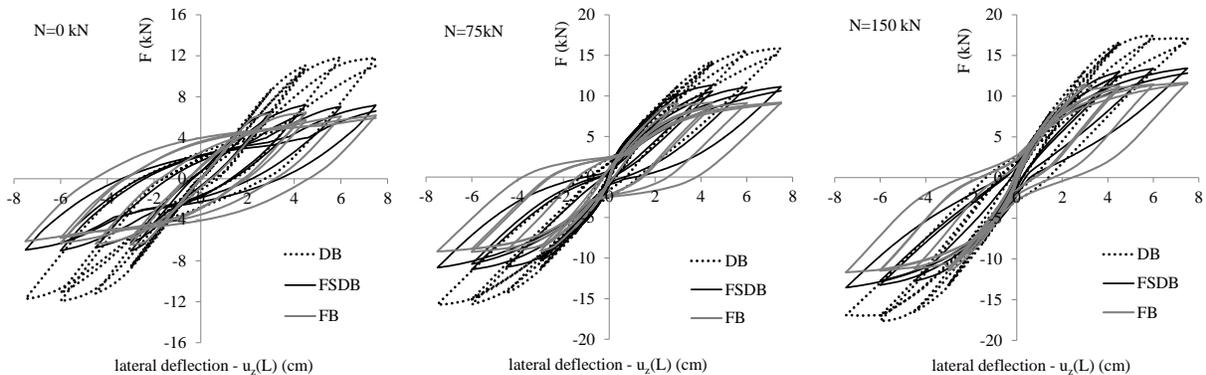

**Fig. 14** Shear force F versus the lateral deflection of the free end of the column $u_z(L)$ **for a cyclic analysis for different levels of axial load.**

**SUMMARY AND CONCLUSIONS**

The shape functions, adopted by classical displacement based finite formulations for the discretisation of the beam element displacement field, are based on the adoption of Hermite polynomials which, however, are not able to capture the curvature variations due to along axis plastic deformation occurrences. The latter circumstance results in the inadequacy of such shape functions to properly represent the displacement field in presence of flexural stiffness variations implied by plastic constitutive behaviour. As a result, a great computational disadvantage is related to the need of adopting refined meshes in order to converge towards more accurate solutions which are, on the contrary, provided by force based finite elements procedures. To make beam finite elements, based on formulation of displacement shape functions, more competitive a fibre smart displacement based (FSDB) beam element has been proposed in this work. The proposed element is defined by displacement shape functions obtained as solution of with stepped variations of axial and flexural stiffness by making use of generalised functions. The relevant shape function are named smart in view of their ability to update the terms dependent on the stiffness steps according to the stiffness decay caused by the plastic deformation occurrences. The stiffness update during the plastic incremental analysis is performed according to a fibre approach very convenient for r/c cross sections in order to account for the interaction of axial force –bending moment. The FSDB beam element is formulated with six degrees of freedom at the element nodes and its smart character provides a better accuracy with respect to classical displacement based beam elements as confirmed by numerical tests with results available in the literature for specific beam-like structures.